\documentclass[twocolumn,showpacs,preprintnumbers,amsmath,amssymb,superscriptaddress]{revtex4}

\usepackage{hyperref}%
\usepackage{graphicx}% Include figure files
\usepackage{dcolumn}% Align table columns on decimal
\usepackage{bm}% bold math
\usepackage{epsf}
\usepackage{mathrsfs}
\begin{document}

\title{Light-driven liquid crystalline nonlinear oscillator under optical periodic forcing}

\author{Dmitry O. Krimer}
\affiliation{Theoretische Physik, Universitaet Tuebingen, 72076
Tuebingen, Germany}
\author{Etienne Brasselet}
\affiliation{Centre de Physique Mol\'eculaire Optique et Hertzienne,
Universit\'e Bordeaux I, CNRS, 351 Cours de la
Lib\'eration, 33405 Talence Cedex, France}

\date{\today}

%%%%%%%%%%%%%%%%%%%%%%%%%%%%%%%%%%%%%%%%%%%%%%%%%%%%%%%%%%%%%%%%%%%%%%%

\begin{abstract}

An all-optically driven strategy to govern a liquid crystalline
collective molecular nonlinear oscillator is discussed. It does not
require external feedbacks of any kind while the oscillator and a
time-depending perturbation both are sustained by incident light. Various dynamical regimes such as frequency-locked,
quasiperiodic, forced and chaotic are observed in agreement with a
theoretical approach developed in the limit of the plane wave approximation.

\end{abstract}

 \maketitle

%%%%%%%%%%%%%%%%%%%%%%%%%%%%%%%%%%%%%%%%%%%%%%%%%%%%%%%%%%%%%%%%%%%%%%%%%

The response of nonlinear oscillators to time-dependent external control
parameter is well-known and has been investigated in a wide range of
systems including physical \cite{Pikovsky01_book}, chemical
\cite{Kuramoto84_book} and biological ones \cite{Winfree00_book}.
Periodic forcing may lead to entrainment or quasiperiodicity according to
whether the ratio $f_F/f_N$ between the forcing frequency ($f_F$) and the
natural frequency of the autonomous system ($f_N$) is rational or
irrational. The natural limit cycle behavior of the oscillator may also
be driven into more complex dynamics. Recent applications of periodic
forcing such as the control of a chaotic chemical reaction
\cite{Cordoba06} or the control of a microfluidic droplet emitter
\cite{Willaime06} have emphasized the use of this technique to engineer
controllable systems.

An optical forcing scheme to achieve a non-contact control of a nonlinear
oscillatory system is limited to a few number of systems. A well-known
example is the light-sensitive Belousov-Zhabotinsky reaction, which is a
spatially extended system where a homogeneous time-periodic optical
forcing allows to control pattern formation \cite{Lin04}. A liquid
crystal light valve, where a two-dimensional external feedback and an
additional quasistatic electric field have been used, is another example
\cite{Gutlich05}. In contrast, a local optical periodic forcing on
uniformly oscillating catalytic surface reaction has been investigated in
Ref.~\cite{Wolff03}, where the heating provided by a focused laser beam
acts as the local external perturbation. To our knowledge the all-optical
situation, where both the nonlinear oscillator and the time-dependent
perturbation are supplied by light only, without need of an external
feedback, has not been reported yet.

We propose an all-optical periodic forcing strategy to control a liquid
crystalline molecular oscillator driven by light based on optical
orientational nonlinearities of mesophases \cite{Tabiryan86_review}.
There is no need of external feedback owing to the strong light-matter
coupling that occurs during the propagation of light through the
optically anisotropic liquid crystal (LC) material. Indeed it is known
that a LC under an intense laser field can be viewed as a collective
molecular oscillator whose dynamics depends on the light-matter
interaction geometry. Very rich director dynamics has been observed and confirmed theoretically at fixed light intensity.
Among most studied geometries one can mention a circularly polarized
beam at normal incidence \cite{Santamato86,Brasselet05},  an elliptically
polarized beam at normal incidence \cite{Vella02,Krimer05_2}, an ordinary
linearly polarized beam at oblique incidence \cite{Cipparrone93} or a
linearly polarized beam at normal incidence having an elliptic intensity
profile \cite{Piccirillo01}. More specifically, laser-induced nonlinear reorientation dynamics in LCs has retained some attention in the context of transition to chaos \cite{Demeter99,Carbone01,Demeter05} and chaotic rotations \cite{Vella03,Brasselet06}.
In contrast to all previous
studies performed at fixed intensity, the aim of the present work is to
explore both theoretically and experimentally two representative light-LC interaction
geometries when the intensity is periodically modulated.
%%%%%%%%%%%%%%%%%%%%%%%%%%%%%%%
\begin{figure}[!b]
\begin{center}
\includegraphics[width=8cm]{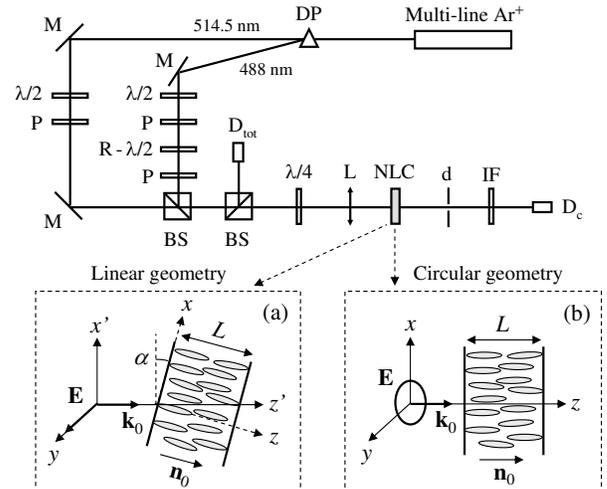}
\end{center}
\caption{Experimental set-up. Multi-line Ar$^+$: Argon-ion laser; DP:
dispersion prism; M: mirror; $\lambda/2$: half-wave plate; P: polarizer;
R-$\lambda/2$: rotating half-wave plate; BS: beam splitter; $\lambda/4$:
quarter-wave plate; L: lens; NLC: film of nematic LC; d: diaphragm; IF:
interferential filter for 514.5~nm; D$_i$: photodiodes. (a,b) linear and
circular geometries; ${\bf k}_0$ and ${\bf E}$: incident wavevector and
electric field.} \label{fig_setup}
\end{figure}
%%%%%%%%%%%%%%%%%%%%%%%%%%%%%%%

The observations are performed using the experimental set-up shown in
Fig.~\ref{fig_setup} where the linear geometry refers to a linearly
polarized light impinging at oblique incidence (typically few degrees)
onto nematic film with the electric field perpendicular to the incidence
plane [Fig.~\ref{fig_setup}(a)]. The circular geometry corresponds to a
circularly polarized light with normal incidence
[Fig.~\ref{fig_setup}(b)]. In both cases the autonomous system has a
limit cycle behavior at the onset of a secondary Hopf bifurcation, where
the reorientation amplitude $|{\bf n}_\perp|$ (${\bf n}_\perp={\bf
n}-{\bf n}_0$ and ${\bf n}_0$ is the unperturbed director, which is the
unit vector that represents the averaged local molecular orientation, see
Fig.~\ref{fig_setup}) acquires an oscillatory behavior with natural
frequency $f_N$. The periodic forcing is then achieved by sinusoidal
perturbation of the intensity, $\rho(t)=\rho_0+\delta\rho\,\sin(2\pi f_F
t)$. Here, $\rho$ is the intensity normalized to the reorientation
threshold, which corresponds to the so-called optical Fr\'eedericksz
transition, and $\delta\rho$ is the forcing amplitude.

We used in the experiment commercial nematic E7 from Merck which was
placed between two glass substrates that were chemically treated to
ensure homeotropic anchoring (molecules are perpendicular to the film
walls). The incident beam is focused onto a nematic film of $L=75~\mu$m
thickness using a 150~mm focal length lens. The aforementioned general
form of $\rho(t)$ was obtained by combining two laser lines
($\lambda_1=514.5$~nm and $\lambda_2=488$~nm) selected from a multi-line
linearly polarized Argon-ion laser using a dispersion prism placed at the
laser output. The total intensity of each of the two beams is controlled
using independent combinations of a $\lambda/2$ plate followed by a
polarizer. The linear polarization of the $\lambda_2$-beam is
continuously rotated owing to a rotating $\lambda/2$ plate controlled by
an electrical motor. The polarizer placed after the rotating $\lambda/2$
plate ensures that $\lambda_2$-beam is polarized along the $y$ direction
with the intensity $\rho_2(t) \propto \sin^2(\pi f_F t)$ whereas the
intensity $\rho_1$ of the $y$-polarized $\lambda_1$ beam is kept fixed.
Both beams then recombine through a beam splitter to generate an
excitation light field whose total intensity $\rho(t)=\rho_1+\rho_2(t)$
is of the required form. Note that the forcing frequency ($f_F$) and
amplitude ($\delta\rho$) can be adjusted independently. Finally, a
$\lambda/4$ plate placed before the lens allows to switch between two
geometries. Its optical axis is set along (at $45^\circ$ of) the $y$-axis
in the linear (circular) case.

\begin{figure}[!b]
\begin{center}
\includegraphics[width=8cm]{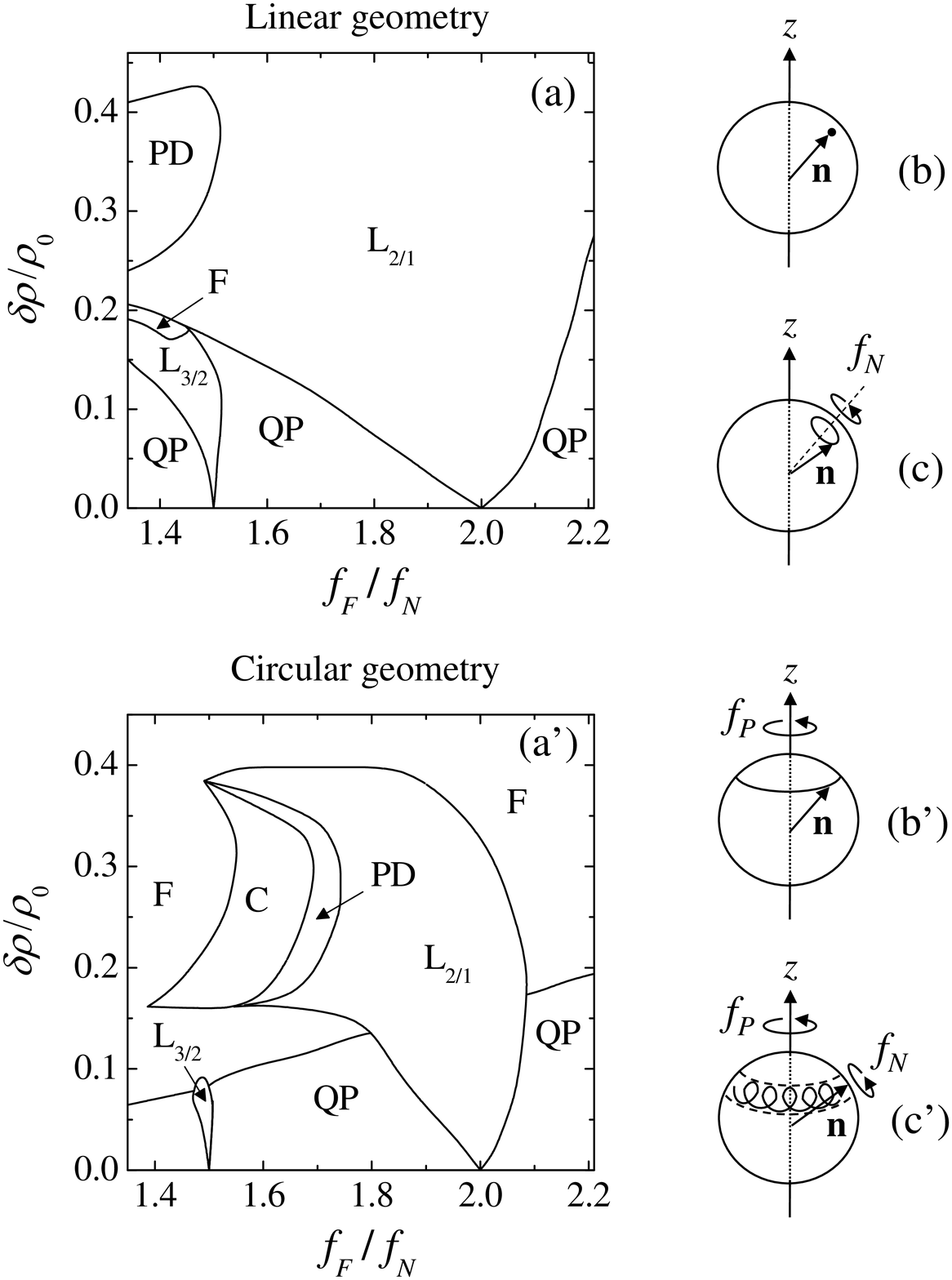}
\end{center}
\caption{Maps of the dynamical regimes near 2:1 entrainment region in linear and
circular geometry. (a,a') L$_{p/q}$: frequency-locked ($f=f_F/p$); F:
forced ($f=f_F$); QP: quasiperiodic; PD: region of period doubling
[$f=f_F/(2n)$, $n=2,3...$]; C: chaos. The director dynamics on the unit
sphere is sketched below (b,b') and above (c,c') the secondary
Hopf bifurcation.}
\label{fig_maps}
\end{figure}

At first the total intensity, which is monitored by the photodiode
$D_{\rm tot}$ (see Fig.~\ref{fig_setup}), is increased smoothly from
zero, setting $\delta\rho=0$. The homeotropic state remains stable below
the Fr\'eedericksz threshold. Above threshold the system settles either
to a stationary distorted state (fixed point) in the linear case or to a
state of uniform precession of the director around the $z$-axis with the
frequency $f_P$ (limit cycle) in the circular case as shown in
Figs.~\ref{fig_maps}(b,b'). A further increase of the intensity $\rho_0$
leads to a secondary supercritical Hopf bifurcation in both cases. At the
onset of this instability, a new frequency $f_N$ associated with the
oscillation of the reorientation amplitude appears as sketched in
Fig.~\ref{fig_maps}(c,c'). Forcing experiments are performed
slightly above the secondary threshold (typically a few percent) ensuring
that a system never reaches higher instabilities for both linear
\cite{Demeter01} and circular geometry \cite{Brasselet05}.

Both the natural limit
cycle at $\delta\rho=0$ and the reorientation amplitude dynamics at
$\delta\rho \neq 0$ are monitored by the time-dependent total intensity
of the central part of the beam that emerges from the sample, $I_{\rm
c}^{\rm tot}(t)=I_{\rm c}^{(\lambda_2)}(t)+I_{\rm c}^{(\lambda_2)}(t)$,
where the contribution $I_{\rm c}^{(\lambda_i)}$ depends on $\rho_i$ and
$|{\bf n}_\perp|$. Indeed $I_{\rm c}^{(\lambda_i)}\propto\rho_i$ at fixed
$|{\bf n}_\perp|$. In addition, the larger $|{\bf n}_\perp|$ is, the
stronger self-focusing effects are and $I_{\rm c}$ is small when $\rho_i$
is fixed. Therefore the Fourier spectrum of $I_{\rm c}^{(\lambda_1)}(t)$
can safely be associated with the one of $|{\bf n}_\perp|(t)$ since
$\rho_1$ is constant. This is achieved by using an interferential filter
operating at $\lambda_1$ to get rid of the $\lambda_2$ contribution where
$\rho_2(t)$ is modulated with frequency $f_F$. The photodiode ${\rm D_c}$
(see Fig.~\ref{fig_setup}) thus collects the signal $I_{\rm
c}^{(\lambda_1)}$ that will be further denoted as $I_{\rm c}$. We notice
that we were able to fully characterize the director dynamics (i.e. its
polar and azimuthal degrees of freedom) using a polarimetric analysis of
the output beam, which is not shown in Fig.~\ref{fig_setup}. However,
only the polar degree of freedom is of interest in the presented results.

%%%%%%%%%%%%%%%%%%%%%%%%%%%%%%%%%%%%%%%%%%%%%%%%%%%%%%%%%%%%%%
%%%%%%%%%%%%%%%%%%%%%%%%%%%%%%%%%%%%%%%%%%%%%%%%%%%%%%%%%%%%%%

A rigorous description of the dynamical properties of nematics is
well-established in a full hydrodynamic approach \cite{deGennes93_book}.
The basic equations are those of the director and the velocity field,
which are coupled with the Maxwell's equations that governs the
propagation of light. The director equations are derived using the torque
balance condition among the elastic, electric and viscous torques. The
flow caused by the director reorientation was shown to lead to
quantitative rather than to qualitative changes in both geometries
\cite{Krimer05,Demeter05} within the range of intensity that has been explored
here. Thus, the velocity is neglected in our study. In the calculations
we used the plane wave approximation and assumed that the director
depends only on $z$ and $t$. Taking into account that $\lambda \ll L$ the
Maxwell's equations were solved under the geometrical optics
approximation. The latter are consequently reduced to a set of two
ordinary differential equations for the amplitudes of the ordinary
($o$) and extraordinary ($e$) waves. Despite the simplification described the
resulting set of equations is rather cumbersome and is not presented here
explicitly. The complete bifurcation scenario for the autonomous system
($\delta\rho=0$) is available for both linear \cite{Demeter01} and
circular geometry \cite{Brasselet05} under the aforementioned
approximations. The models developed there might straightforwardly be
extended to the case of time-dependent intensity $\rho(t)$. As in
previous studies, we used different representations to describe the
director depending on the geometry. Such a choice is dictated by symmetry
considerations. Then an expansion of the director components is performed
in terms of orthogonal functions which satisfy the boundary conditions
$n_{x,y}(z=0,L;t)=0$. Namely we took ${\bf n} = (\sin\theta, \cos\theta
\sin\phi,$ $\cos\theta\cos\phi)$ with $\phi=\sum_{n=1}^{\infty}
\phi_n(t) \, \sin(n\pi z/L)$ and $\theta=\sum_{n=1}^{\infty} \theta_n(t) \,
\sin(n\pi z/L)$ for linear geometry, whereas ${\bf n} = (\sin\Theta
\cos\Phi, \sin \Theta\sin\Phi,\cos\Theta)$ with
$\Theta=\sum_{n=1}^{\infty} \Theta_n(t) \, \sin(n\pi z/L)$ and
$\Phi=\Phi_0(t)+\sum_{n=1}^{\infty}\Phi_n(t) \, \sin[(n+1)\pi
z/L]/\sin(\pi z/L)$ for the circular one. The Galerkin procedure was then
used to obtain a set of nonlinear ordinary differential equations for the mode amplitudes
$(\phi_n,\theta_m)$ or $(\Theta_n,\Phi_m)$ which was solved numerically using
standard Runge-Kutta method. In fact it is enough to retain only a few
number of modes for the director expansion to obtain a good accuracy for
the calculated director components (better than $1\%$). Note that the
Maxwell's equations were solved at each step of numerical integration for
time $t$. We also introduce the total phase delay $\Delta(t) =
2\pi/L\int_0^L[n_e(z,t)-n_o]\,dz$ between the $o$- and $e$-waves across
the whole film ($n_{e,o}$ are the refractive indices), which is a global
measure of the amplitude of reorientation \cite{Tabiryan86_review}. The
calculated $\Delta(t)$ will thus be compared with the measured $I_{\rm
c}(t)$.
%
%%%%%%%%%%%%%%%%%%%%%%%%%%%%%%%%%%%%%%%%%%%%%%%%%%%%%%%%%%%%%%
\begin{figure}[!ht]
\begin{center}
\rotatebox{0}{\includegraphics[width=8.5cm]{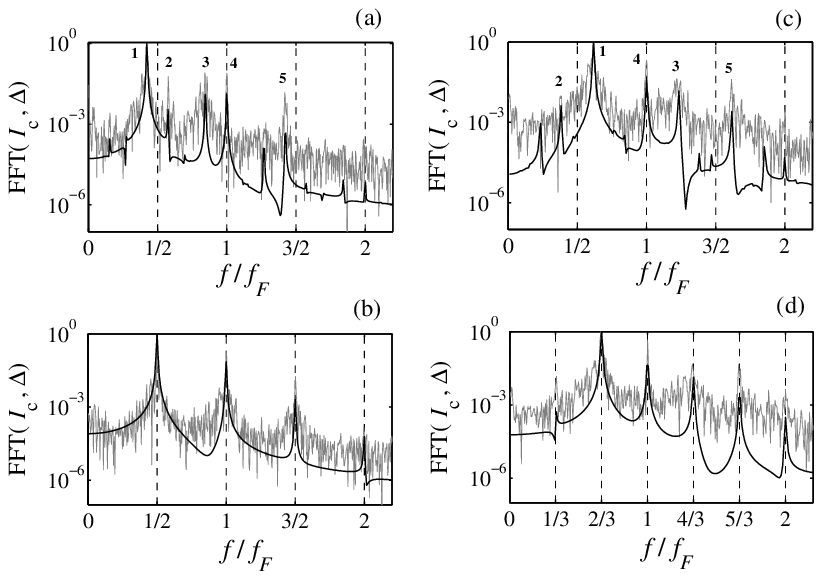}}
\end{center}
\caption{Demonstration of QP, L$_{2/1}$ and L$_{3/2}$ regimes in linear
geometry for $\delta\rho/\rho_0=0.05$ and $\alpha=3.5^\circ$. Gray (black)
curves: experimental (calculated) power spectra of $I_{\rm c}$
($\Delta$). (a) right part of the L$_{2/1}$ tongue at $f_F/f_N=2.37$; (b)
inside the L$_{2/1}$ tongue at $f_F/f_N=2$; (c) left part of the
L$_{2/1}$ tongue at $f_F/f_N=1.62$;  (d) inside the L$_{3/2}$ tongue at
$f_F/f_N=1.5$. For QP states (a) and (c) the frequency peaks (1-5) refer
to ($f_N$, $f_F-f_N$, $2f_N$, $f_F$, $f_N+f_F$) respectively.}
\label{fig_expQPL}
%\end{figure}
%%%%%%%%%%%%%%%%%%%%%%%%%%%%%%%%%%%%%%%%%%%%%%%%%%%%%%%%%%%%%%
%%%%%%%%%%%%%%%%%%%%%%%%%%%%%%%%%%%%%%%%%%%%%%%%%%%%%%%%%%%%%%
%\begin{figure}[!ht]
\begin{center}
\rotatebox{0}{\includegraphics[width=8.5cm]{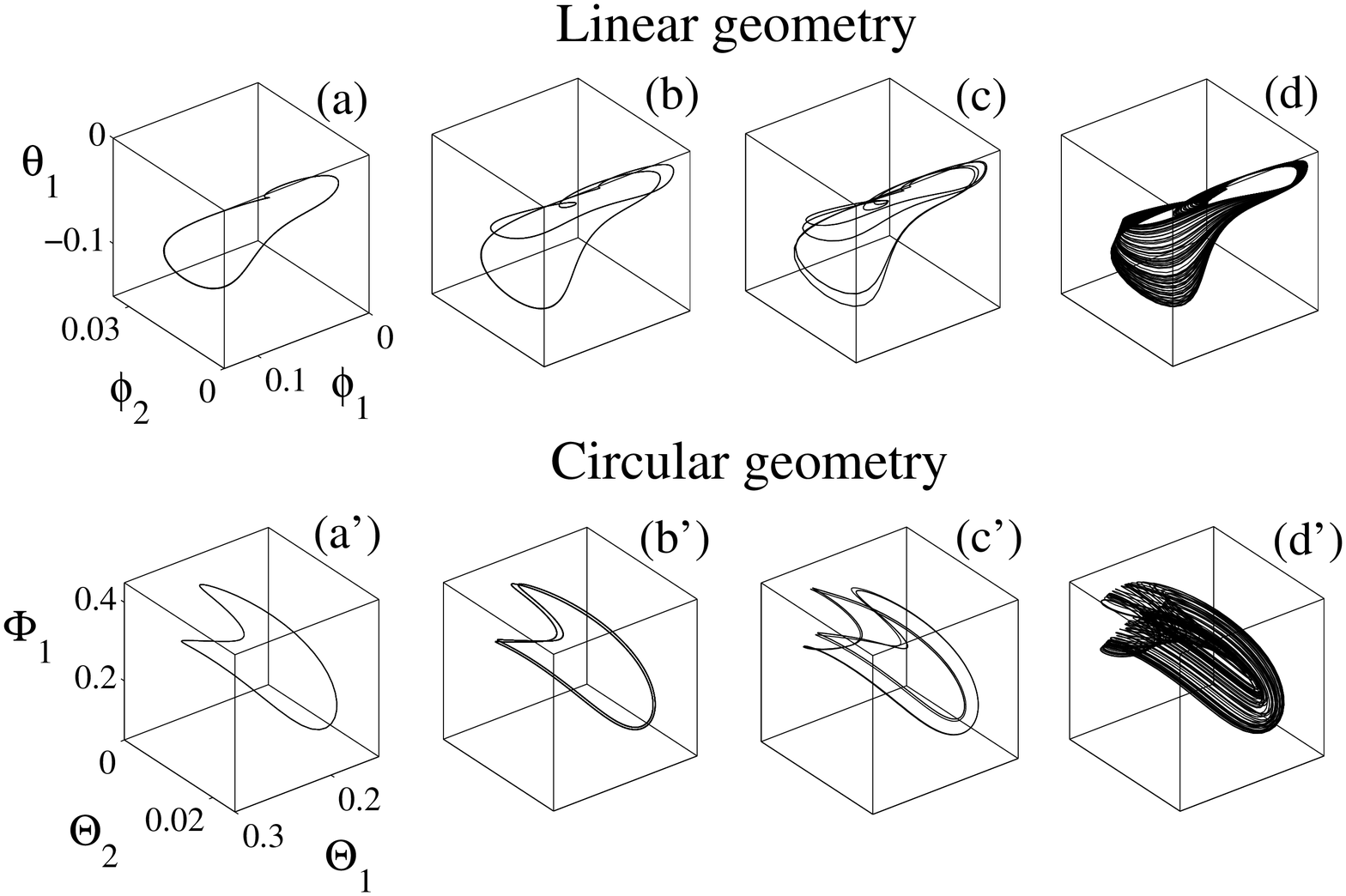}}
\end{center}
\caption{Director trajectories in the ($\phi_1, \phi_2, \theta_1$) [($\Theta_1,
\Theta_2, \Phi_1$)] phase space for linear [circular] geometry
illustrating the transition to chaos at $\delta\rho/\rho_0=0.3$
[$\delta\rho/\rho_0=0.27$] for the same parameters as for the maps shown
in Fig.~\ref{fig_maps}. (a,a'): $f_F/f_N=1.5$ [$1.767$] (limit cycle in
the L$_{2/1}$ regime); (b,b'): $f_F/f_N=1.35$ [$1.742$] (limit cycle
after the first period doubling); (c,c'): $f_F/f_N=1.27$ [$1.691$] (limit
cycle after the second period doubling); (d,d'): $f_F/f_N=1.18$ [$1.632$]
(chaotic attractor).} \label{fig_trajectories}
\end{figure}
%%%%%%%%%%%%%%%%%%%%%%%%%%%%%%%%%%%%%%%%%%%%%%%%%%%%%%%%%%%%%%

The calculated maps of dynamical regimes are shown in Fig.~\ref{fig_maps}
in the $(\delta\rho/\rho_0,f_F/f_N)$ plane, where the same material
parameters as in Ref.~\cite{Brasselet05} have been used. In addition,
$\rho_0$ is taken approximately 1\% above the secondary Hopf bifurcation
threshold. For the sake of illustration the results are presented for a
typical layer thickness $L=100~\mu$m and a typical incidence angle
$\alpha=5^\circ$ (for the linear geometry only, see
Fig.~\ref{fig_setup}). These maps aim to show the generic dynamical
behavior of the system and further quantitative comparison with
experiments is done using the actual experimental values. We note however
that the conclusions are unchanged under (i) cell thickness changes as
long as the geometrical optics approximation is satisfied and (ii)
incidence angle changes in the typical range 2-8$^\circ$ [for linear
geometry] since the observed dynamics is mainly dictated by existence of
the secondary Hopf instability \cite{Demeter01}. In both cases we focus
our study on the region near 2:1 resonance which is known to be strong.
An advantage of the system in question is that the entrainment region
(the so-called Arnold's tongue) which emanates from the point $(2/1,0)$
is rather wide, which facilitates its observation, see
Fig.~\ref{fig_maps}. Experimental evidence of entrainment is found by
fixing the forcing amplitude and scanning the forcing frequency. The
results for the linear geometry are shown in Fig.~\ref{fig_expQPL}, where
the power Fourier spectra of $I_{\rm c}$ are plotted together with the
spectra of its theoretical analog, the phase delay $\Delta$. Each panel
corresponds to a measurement which lasted $\sim4000$~s, with $f_F$
decreasing from the panel (a) to (d). The spectra of $I_{\rm c}$ and
$\Delta$ for a state within the 2:1 tongue [Fig.~\ref{fig_expQPL}(b)] are
characterized by a single frequency, $f_F/2$, as expected. Outside the
tongue and for sufficiently small $\delta\rho$, the system responds
quasiperiodically. In that case the spectra of $I_{\rm c}$ and $\Delta$
are described by means of two incommensurable frequencies $f_N$ and $f_F$
that generate peaks at frequencies $nf_N\pm m f_F$, $n$ and $m$ integers,
as shown in Fig.~\ref{fig_expQPL}(a,c). The 3:2 tongue is also observed
and shown in Fig.~\ref{fig_expQPL}(d). In this case the single frequency
$f_F/3$ determines the dynamics, as expected.
%
%%%%%%%%%%%%%%%%%%%%%%%%%%%%%%%%%%%%%%%%%%%%%%%%%%%%%%%%%%%%%%
\begin{figure}[!tbp]
\begin{center}
\rotatebox{0}{\includegraphics[width=8.5cm]{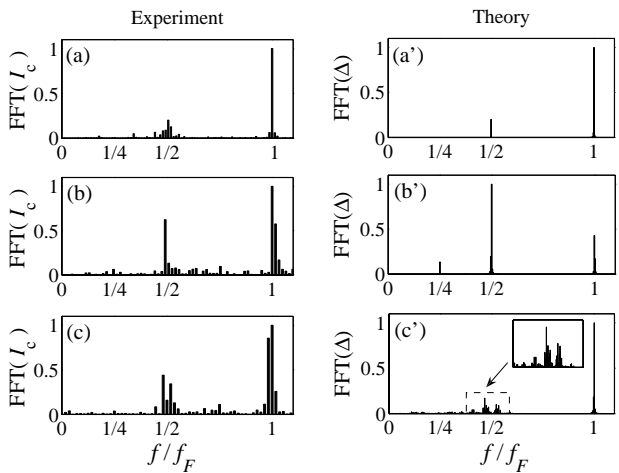}}
\end{center}
\caption{ Demonstration of the sequence L $\to$ PD$\to$C (circular
geometry). Power spectra of $I_{\rm c}$ (experiment) and $\Delta$
(calculated) are shown at $\delta\rho/\rho_0=0.22$. (a,a'): locked state
at $f_F/f_N\simeq2.05$ and $f_F/f_N=2.06$; (b,b'): state within PD region
at $f_F/f_N\simeq1.6$ and $f_F/f_N=1.65$; (c,c'): chaotic state at
$f_F/f_N\simeq1.45$ and $f_F/f_N=1.58$.} \label{fig_expchaos}
\end{figure}
%%%%%%%%%%%%%%%%%%%%%%%%%%%%%%%%%%%%%%%%%%%%%%%%%%%%%%%%%%%%%%
%
%%%%%%%%%%%%%%%%%%%%%%%%%%%%%%%%%%%%%%%%%%%%%%%%%%%%%%%%%%%%%
\begin{figure}[!ht]
\begin{center}
\rotatebox{0}{\includegraphics[width=7cm]{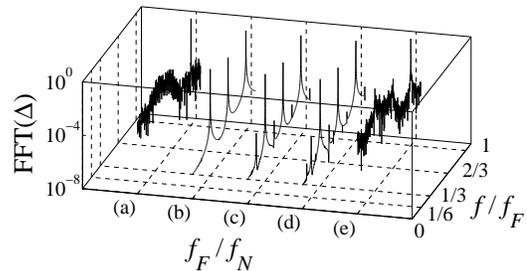}}
\end{center}
\caption{Window of regular dynamics in the chaotic region for
$\delta\rho/\rho_0=0.22$ in the circular case. (a) $f_F/f_N=1.525$
[chaotic state]; (b) $f_F/f_N=1.5195$ [periodic state with $f=f_F/3$];
(c) $f_F/f_N=1.5184$ [periodic state with $f=f_F/6$]; (d)
$f_F/f_N=1.5182$ [periodic state with $f=f_F/12$]; (e) $f_F/f_N=1.5165$
[chaotic state].} \label{fig_regularwindow}
\end{figure}
%%%%%%%%%%%%%%%%%%%%%%%%%%%%%%%%%%%%%%%%%%%%%%%%%%%%%%%%%%%%%
%
For higher forcing amplitude a forced regime (the F region in
Fig.~\ref{fig_maps}) was found which has $f_F$ as a unique characteristic
frequency. When starting inside the 2:1 tongue and decreasing the forcing
frequency (keeping the same value of $\delta\rho$), a route to chaos via
a cascade of period-doubling bifurcation was found for moderate to large
forcing amplitudes (the PD and C regions in Fig.~\ref{fig_maps}). The
resulting typical trajectories are shown in Fig.~\ref{fig_trajectories} for both linear
(at $\delta\rho/\rho_0=0.3$) and circular (at $\delta\rho/\rho_0=0.27$)
geometries. In Fig.~\ref{fig_expchaos} the power spectra of $I_{\rm c}$
(experiment) and $\Delta$ (theory) are compared for the successive states
from the sequence L $\to$ PD $\to$ C for the circular geometry. In
particular the transition to a chaotic regime is accompanied by an abrupt
frequency widening around $f/f_F=1/2$. Such a behavior is predicted by
theory as well [see Figs.~\ref{fig_expchaos}(b',c')]. In addition, the
characteristic presence of windows of regularity inside the chaotic
region when $f_F$ varies at fixed $\delta\rho$ has been found. There, the
director dynamics is periodic. One of these windows is shown in Fig.~\ref
{fig_regularwindow} for $\delta\rho/\rho_0=0.22$, where a $f_F/3$
dynamics suddenly appears when the frequency is decreased. Period
doubling then begins again with limit cycles having frequencies $f_F/6$,
$f_F/12$, etc. and then once again break off to chaos.

In conclusion, we showed that an all-optical control scheme based on
periodic forcing of the light intensity can be implemented in liquid
crystalline materials. Two light-matter interaction geometries have been
used for demonstration. Various dynamical regimes such as
frequency-locked, quasiperiodic, forced or chaotic have been shown to
result from the optical forcing. The theoretical study is carried out
within the general framework involving Maxwell's equations and the
constitutive material equations. Fairly good qualitative agreement between theory and experiments is obtained in the entire range of
the problem parameters. The polarization sensitivity of mesophases might
also offer the possibility to extend present results to a periodic
modulation of the polarization instead of the intensity.

This paper is dedicated to the memory of Prof. Lorenz Kramer who suddenly
passed away on 05/04/2005. The preliminary idea of that work was discussed with him. We thank Prof. Michael Tribelsky
for his valuable comments. EB thanks the Physics Laboratory of Ecole Normale Sup\'erieure de Lyon in France for the use of its experimental facilities.

%%%%%%%%%%%%%%%%%%%%%%%%%%%%%%%%%%%%%%%%%%%%%%%%%%%%%%%%%%%%%
% BIBLIOGRAPHY
%%%%%%%%%%%%%%%%%%%%%%%%%%%%%%%%%%%%%%%%%%%%%%%%%%%%%%%%%%%%%

\end{document}